\documentclass[review]{elsarticle}
\usepackage{graphicx}
\usepackage{hyperref}
\usepackage{color}
\usepackage{amsmath}
\usepackage{amsfonts}
\usepackage{amssymb}
\usepackage{booktabs}
\usepackage{threeparttable}
\usepackage{multirow}
\usepackage{tabulary}
\usepackage{acronym}
\usepackage{xspace}
\usepackage[nameinlink,capitalize,noabbrev]{cleveref}
\usepackage{hyphenat}
\makeatletter

\NewDocumentCommand{\vmwe}{}{5\,MWe\xspace}

\acrodef{MCMC}{Markov chain Mone Carlo}
\acrodef{BU}{Burnup}
\acrodef{CT}{Cooling time}

 \usepackage{float}
\begin{document}
\title{Reconstructing North Korea's Plutonium Production History with Bayesian Inference-Based Reprocessing Waste Analysis}
\author[tuda,rwth]{Benjamin Jung\corref{cor1}}
\ead{benjamin.jung1@rwth-aachen.de}
\author[rwth]{Johannes Bosse}
\author[prif,tuda]{Malte Göttsche}
\cortext[cor1]{Corresponding author}
\affiliation[tuda]{
    organization={Technical University of Darmstadt},
    addressline={Karolinenplatz 5},
    city={Darmstadt},
    postcode={64289},
country={Germany}
}
\affiliation[rwth]{organization={Nucler Verification and Disarmament, RWTH Aachen University},addressline={Schinkelstrasse 2},
    city={Aachen},
    postcode={52062},
country={Germany}}
\affiliation[prif]{
    organization={Peace Research Institute Frankfurt},
    addressline={Baseler Str. 27--31},
    city={Frankfurt am Main},
    postcode={60329},
country={Germany}
} \begin{abstract}

Although North Korea's nuclear program has been the subject of extensive scrutiny, estimates of its fissile material stockpiles remain fraught with uncertainty.
In potential future disarmament agreements, inspectors may need to use nuclear archaeology methods to verify or gain confidence in a North Korean fissile material declaration.
This study explores the potential utility of a Bayesian inference-based analysis of the isotopic composition of reprocessing waste to reconstruct the operating history of the \vmwe reactor and estimate its plutonium production history.
We simulate several scenarios that reflect different assumptions and varying levels of prior knowledge about the reactor.
The results show that correct prior assumptions can be confirmed and incorrect prior information (or a false declaration) can be detected.
Model comparison techniques can distinguish between scenarios with different numbers of core discharges, a capability that could provide important insights into the early stages of operation of the 5 MWe reactor.
Using these techniques, a weighted plutonium estimate can be calculated, even in cases where the number of core discharges is not known with certainty.

\end{abstract}  \maketitle

\hypertarget{introduction}{\section{Introduction}\label{introduction}}
Fissile material accounting is one of the cornerstones of the international non-proliferation regime and is likely to be equally instrumental to future nuclear disarmament agreements.
Verifying a declared fissile material inventory is difficult, especially if the past operation of nuclear facilities has previously been concealed from the international community.
To build confidence in a fissile material declaration, it will be necessary to reconstruct the operating histories of these facilities.
To support such confidence-building measures, nuclear archaeology offers a useful set of tools, including forensic measurements of samples from structural reactor components to reconstruct the operational history \cite{gottscheNuclearArchaeologyAssess2024}.

In a recent study \cite{JUNG20242704}, we presented a promising new nuclear archaeology concept that uses a Bayesian inference framework to reconstruct reactor operating parameters from samples of nuclear reprocessing waste.
We call this framework the Bayesian Reprocessing Waste Analysis Method (BRAM).
The results indicated that it is in principle possible to reconstruct the average fuel burnup (BU) and the cooling time (CT) of the high-level waste (HLW) of two different reactor cycles, even if the waste has been mixed.
However, the parameter reconstruction was not always successful if the average fuel burnup was low ($\leq 3 \text{\,MWd/kg}$).

This study explores the applicability of this Bayesian inference framework to a hypothetical future scenario in which the Democratic People's Republic of Korea (DPRK) has declared its fissile material inventory and inspectors are tasked with verifying the declaration.
North Korea has operated its \vmwe reactor for at least five cycles, although there are doubts about the number of full-core discharges.
The average fuel burnup of each (possibly) discharged core has been estimated to be below 3\,MWd/kg \cite{Kütt}.
Therefore, the scenario is useful to test and develop the concepts of the framework and to investigate its potential utility for verifying a future declaration of the fissile material inventory of the DPRK.

This study pursues several research goals.
First, to test BRAM on scenarios where waste from more than two (i.e., five or six) reactor cycles has been mixed.
Second, to investigate if the method can be used to assess whether the measured samples are consistent with a declared reactor operation history.
Third, to explore the use of model selection techniques to compare two different, plausible reactor histories.

To this end, we simulate two operating histories of the \vmwe reactor and use the Bayesian framework to reconstruct the operating parameters in different scenarios.
 \hypertarget{background}{\section{Background}\label{background}}

It was not until the 1970s that North Korea began to develop a nuclear programme in earnest, although North Korea had been actively pursuing nuclear research programmes since the 1950s.
In 1974, North Korea became a member of the International Atomic Energy Agency (IAEA), joined the NPT in 1985 and finished building its 5\,MWe experimental reactor in Yongbyon by 1986.
It uses natural uranium as fuel, graphite as neutron moderator, and is cooled with CO$_2$. This makes it well suited for the production of plutonium for nuclear weapons.
Since the 1990s, the United States and North Korea have been in negotiations over North Korea's nuclear program \cite{CLS}.
In 1992, the IAEA found inconsistencies in North Korea's declaration of its nuclear material with regard to correctness and completeness.
This assessment was based on swipe samples at declared nuclear facilities, samples from North Korea's declared plutonium and from waste tanks containing trace amounts of plutonium \cite{Albright}.

Until today, there remains uncertainty about the operational history of the 5\,MWe reactor within the time span of 1986-1994.
The reactor was shut down for the first time in 1989 for a period of 70 to 100 days.
It is possible that the DPRK may have discharged either a partial or full core at that time.
In 1994, the reactor was shut down and defueled.
Later that year, North Korea and the United States signed the Agreed Framework, leading North Korea to halt its production of fissile material for nine years.
In 2003, the reactor was restarted, and North Korea reprocessed the irradiated fuel that had been unloaded in 1994.
The reactor then operated during two production campaigns: from February 2003 to April 2005, and from June 2005 to July 2007.
The fuel irradiated during both campaigns was subsequently reprocessed.
In June 2008, as part of the “six-party talks” on the denuclearization of the Korean peninsula, North Korea destroyed the reactor’s cooling tower as a gesture of goodwill.
However, the reactor resumed operations from August 2013 to October 2015, with the irradiated fuel reprocessed the following year.
In early 2016, the reactor was started again and continued running until spring 2018, with the irradiated fuel reprocessed shortly thereafter.
There is no evidence that the reactor operated in 2019 and 2020 \cite{Kütt}.

\begin{table}[tbh]
  \centering
  \begin{threeparttable}
      \caption[Two potential operating histories of the 5\,MWe reactor in Yongbyon]{
        Two potential operating histories of the 5\,MWe reactor in Yongbyon.
        The operating windows and estimated average burnup values are based on \cite{Kütt}.
        }
      \label{tab_reactor_history}
      \begin{tabular}{ccc}
          \toprule
         Operating window  & \multicolumn{2}{c}{Burnup [MWd/kg]}\\
           & Pre-1994 disch. & 1994 disch. \\
         \midrule
         1986--1989 & 0.185 & - \\
         1989--1994 & 0.5--0.7 & - \\
          1986--1994  & - & 0.6--0.7 \\
           Jan.\,2003--Apr.\,2005 & 0.23--0.33 & 0.23--0.33 \\
           Jun.\,2005--Jul.\,2007 & 0.22--0.31 & 0.22--0.31 \\
           Aug.\,2013--Oct.\,2015 & 0.22--0.31 & 0.22--0.31\\
           Jan.\,2016--Mar.\,2018 & 0.23--0.32 & 0.23--0.32\\
           \bottomrule
      \end{tabular}
  \end{threeparttable}
  \end{table}

For this paper, we use the two potential reactor histories discussed by \citet{Kütt} and do not consider any potential reactor activities after 2020.
They are shown in \cref{tab_reactor_history}.
Each operating window corresponds to a full core being irradiated in and subsequently discharged from the reactor.
The ``pre-1994 discharge'' scenario assumes that a full core discharge occurred in 1989, whereas the ``1994 discharge'' scenario assumes that the first core remained in the reactor until 1994.
We do not consider any partial discharge scenarios.

In these scenarios, each core is fully reprocessed at some point in time after its discharge and the waste from this process is added to a single waste tank.
We refer to the waste from a single core as a ``batch'', and thus a hypothetical sample of waste from the waste tank would contain a mixture of five or six batches, depending on the scenario.
 \hypertarget{methodology}{\section{Methodology}\label{methodology}}

\begin{figure}
    \centering
    \includegraphics{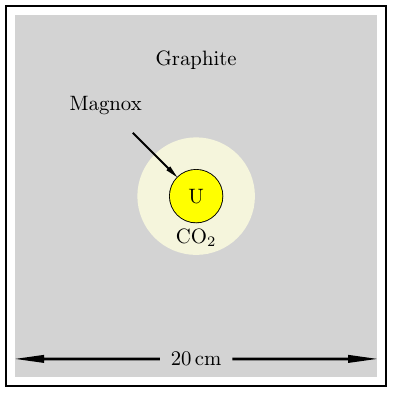}
    \caption{Schematic of a fuel channel of the \vmwe reactor as modelled with \texttt{OpenMC} \cite{romanoOpenMCStateoftheartMonte2015} and \texttt{ONIX} \cite{detroulliouddelanversinONIXOpensourceDepletion2021}.}
    \label{fig:reactor_model}
\end{figure}

\begin{table*}[tbh]
    \centering
        \caption[Design parameters of the \vmwe reactor]{
            Design parameters of the \vmwe reactor.
            The values are based on \citet{Albright} and \citet{Kütt}.
        }
        \label{tab:5mwe-params}
        \begin{tabular}{lclc}
            \toprule
            Fuel composition & nat. U 0.5\,\%-at Al & Moderator material & Graphite \\
            Fuel diameter & 2.9\,cm & Moderator density & 1.65\,g\,$\text{cm}^{-3}$ \\
            Fuel density  & 18.17\,g\,$\text{cm}^{-3}$ & Number of fuel channels & $\approx 800$ \\
            Cladding composition & Mg 1\,\%-at Al & Total fuel mass & $\approx 50\,\text{t}$\\
            Cladding thickness & 0.5\,cm & Thermal power & 20--25 MWt\\
            Channel diameter & 6.5\,cm & Lattice pitch & 20\,cm \\
            \bottomrule
        \end{tabular}
\end{table*}

\subsection{Bayesian Inference}

To reconstruct the operating parameters of each reactor cycle in the \vmwe history, we use the Bayesian inference framework developed in \cite{Figueroa} and \cite{JUNG20242704}.
The purpose of this framework is to reconstruct the posterior distribution of the variables of interest, i.e., the parameters \emph{burnup} $BU$ and \emph{cooling time} $CT$, conditional on observed isotopic ratios $\vec{r}_{obs}$ in the reprocessing waste sample.

Generally, the posterior distribution $p(\vec{\theta}|\vec{r}_{obs})$  of the parameters $\vec{\theta}$ is proportional to the product of the \emph{likelihood} $p(\vec{r}_{obs}|\vec{\theta})$ and the \emph{prior} $p(\vec{\theta})$:

\begin{equation}
    p(\vec{\theta}|\vec{r}_{obs}) \propto p(\vec{r}_{obs}|\vec{\theta}) \cdot p(\vec{\theta}).
    \label{eq:bayes}
\end{equation}
Provided with a computational model for the likelihood, the posterior distribution can be approximated well with \ac{MCMC} methods.
We use the \texttt{PyMC} \cite{salvatierProbabilisticProgrammingPython2016} library to implement the inference models and perform the \ac{MCMC} sampling.

\subsection{The likelihood function}\label{subsec:likelihood}

The likelihood of the observed isotopic ratios $\vec{r}_{obs}$ given values of the inference parameters $\theta$ is modelled as a multivariate normal distribution.
By approximating the isotopic ratios as independent, the likelihood factorizes:
\begin{equation}
    p(\vec{r}_{obs}|\vec{\theta}) = \prod_{i=1}^{N} \mathcal{N}\left(r_{obs,i} - M_i(\vec{\theta}),\sigma_{r,i} \right).
    \label{eq_05_likelihood}
\end{equation}
The index $i$ runs over the observed isotopic ratios and the model $M_i$ is a function of the inference parameters that computes the corresponding isotopic ratio.
Each ratio is associated with an uncertainty $\sigma_{r,i}$, which is intended to capture both measurement and modelling uncertainties.
Since this study uses simulated isotopic ratio values in lieu of experimental mass spectrometry measurements, we approximate $\sigma_{r,i}$ as $5\,\%$ of the value of $r_{obs,i}$.

\subsubsection*{Modelling the composition of the \vmwe reprocessing waste}

To model the isotopic composition of the mixed reprocessing waste sample, we compute the nuclide densities of each isotope in each of the discharged cores, as a function of the cycle parameters $\theta_c = (\text{\ac{BU}}_c,\text{\ac{CT}}_c)$.
Using $F_j(\theta_c)$ to represent the model of the nuclide $j$ after one reactor cycle $c$, the model $M_i$ from \cref{eq_05_likelihood} can be written as:
\begin{equation}\label{eq:ratio-model}
    M_i = \frac{\sum_c^{N_{cycles}}F_{j}(\theta_c)}{\sum_c^{N_{cycles}}F_{k}(\theta_c)}.
\end{equation}
Here, we neglect to model mixing ratios since the fuel volume of each discharged core is the same.

As in our previous work \cite{JUNG20242704}, we use Gaussian process-based surrogate models to circumvent the high computational cost of Monte Carlo neutron transport and depletion calculations.
The training data for the surrogate models is simulated with infinite lattice calculations of a fuel channel (see \cref{fig:reactor_model} and \cref{tab:5mwe-params}) of the \vmwe reactor, using \texttt{OpenMC} \cite{romanoOpenMCStateoftheartMonte2015} to simulate neutron transport and \texttt{ONIX} \cite{detroulliouddelanversinONIXOpensourceDepletion2021} to perform the depletion calculation.
These simulation tools are used with ENDF/B-VIII cross-section data \cite{brownENDFBVIIITh2018}.

\subsubsection*{Selecting useful isotopic ratios}
Generally, nuclear reprocessing waste contains many different elements and isotopes, providing a basis for a large number of isotopic ratios that can be used to inform the posterior.
However, using all of these ratios as observables is not practical, not only due to the effort required to measure them, but also because the Markov chain Monte Carlo sampling algorithm becomes inefficient and struggles to converge.
Therefore, it is useful to select a subset of isotopic ratios that is especially informative on the parameters of interest.
Such a list of isotopic ratios was derived by \citet{Figueroa} with an algorithm that selects isotopic ratios based on their sensitivity to the parameters of interest.

This list of ratios is not very sensitive in the ``low burnup'' region (see \cite{JUNG20242704}), which is, however, precisely the regions of interest in this study.
Therefore, we use the same algorithm as \citet{Figueroa} to select a new set of isotopic ratios to use in the present scenarios.

The algorithm proceeds as follows:
First, a set of test inputs is sampled from the parameter space, and then test outputs, i.e., isotopic ratios, are computed for each input.
Then, the algorithm iterates over all possible pairs of isotopic ratios and approximates the relative standard deviation of the marginal posterior distributions of the inference parameters.
This step is repeated for each entry in the test set.
For each test point, the isotopic ratio pair with the lowest relative standard deviation is added to the final list of isotopic ratios.
The resulting list of isotopic ratios is shown in \cref{tab:ratio-list}.

It is important to note that in our simulation-based setting, we assume that ICP-MS would typically be used, though we do not assess detailed measurement strategies or associated uncertainties.
In principle, domain knowledge could be used to exclude further isotopes from the selection algorithm or to adjust the uncertainty assigned to each ratio.
Nevertheless, the ratios selected for this study serve to demonstrate the concept of the framework.

\begin{table}
    \centering
    \caption{List of isotopic ratios selected for informing the posterior with the algorithm in \cite{Figueroa}.}
    \label{tab:ratio-list}
    \begin{tabular}{cccc}
        \toprule
        Ba-134/Ba-136 & Ba-134/Ba-137 & Cd-110/Cd-113 \\
        Eu-151/Eu-155 & Eu-153/Eu-155 & Gd-154/Gd-155 \\
        Gd-154/Gd-157 & Sm-148/Sm-149 \\
        \bottomrule
    \end{tabular}
\end{table}

\subsection{The prior distributions}\label{ubsec:prior} 

Generally, the prior distributions might be determined from data collected from operating records, data collected via satellite surveillance, expert assessments, or physical constraints.
In this study, the priors of \ac{BU} and \ac{CT} of each operating cycle are both chosen to be uniform distributions within respective intervals.
To understand the utility of BRAM for verification, we analyze several cases with different prior intervals.

In the first case, the operating windows in \cref{tab_reactor_history} are used to define prior intervals on the cooling time and the estimated burnup intervals define the priors of the burnup parameters.
\citet{Kütt} derived these burnup estimates based on the operating windows and some assumptions about the capacity factor of the reactor.
We call this case \emph{realistic}, as it represents hypothetical inspectors using the currently available information to reconstruct the reactor history.

In a second case, we extend the prior intervals beyond the realistic intervals to further assess the sensitivity of the selected isotopic ratios with respect to the inference parameters.
These priors are called \emph{wide} priors, even though the intervals are still physically reasonable and, in the case of burnup, well below the limit where the reactor loses criticality.

The third case is called \emph{wrong priors} and represents a case where inspectors have incorrect information about the reactor.
In this case, the prior intervals for burnup are shifted upwards, such that the true value does not lie within them, corresponding to potential errors in estimates of the capacity factor or power.
The cooling time intervals remain unchanged.

\subsection{Comparing potential histories}

So far, the Bayesian inference framework has been used to infer reactor operating parameters given a fixed number of waste batches.
In the present application scenario, it is unclear whether the reactor core was discharged before 1994 or not, i.e., the number of batches required in the inference model is unknown.
To resolve this question, we propose to use leave-one-out cross-validation (LOO-CV) to  compare two inference models that reflect these two possibilities.

LOO-CV is a method for estimating the out-of-sample prediction accuracy of a Bayesian model and is commonly used to compare different models fitted to the same data \cite{vehtariPracticalBayesianModel2017}.
In this application, a score is assigned to the likelihood of each isotopic ratio of each of the two models conditioned on the simulated observations.
Using this score as a weight, the inference models can be ranked to select the model that best explains the observations.
We use the LOO-CV algorithm implemented in the \texttt{Python} library \texttt{ArviZ} \cite{arviz_2019} to compute the weights and rank the models.
 \hypertarget{inference-results}{\section{Inference Results }\label{inference-results }}

\begin{figure}
    \centering
    \includegraphics[width=\textwidth]{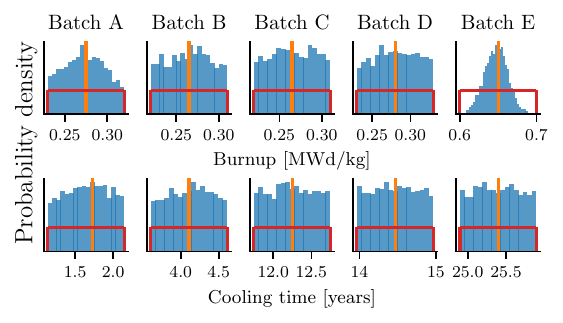}
    \caption{
        Inference results of the test case with \emph{realistic priors}.
        The posterior samples are illustrated by the blue histogram, the vertical orange line indicates the ``true'' parameter value and the red line indicates the respective prior interval.
    }
    \label{fig:1994_scenario}
\end{figure}
\begin{figure}
    \centering
    \includegraphics[width=\textwidth]{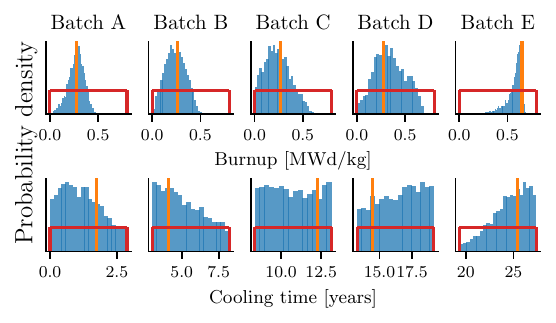}
    \caption{
        Inference results of the test case with \emph{wide priors}.
        The posterior samples are illustrated by the blue histogram, the vertical orange line indicates the ``true'' parameter value and the red line indicates the respective prior interval.
    }
    \label{fig:full_hist_5_batch_wide_priors}
\end{figure}
\begin{figure}
    \centering
    \includegraphics[width=\textwidth]{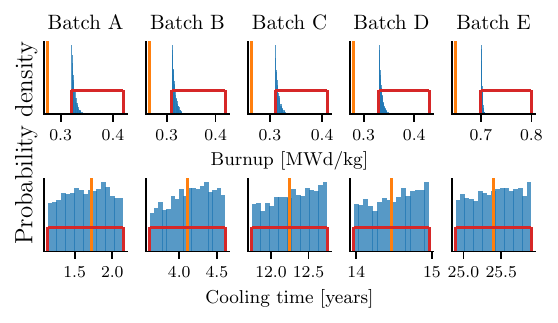}
    \caption{
        Inference results of the test case with \emph{wrong priors}.
        The posterior samples are illustrated by the blue histogram, the vertical orange line indicates the ``true'' parameter value and the red line indicates the respective prior interval.
    }
    \label{fig:full_hist_5_batch_wrong_priors}
\end{figure}

\subsection{The effect of different prior intervals}\label{subsec:prior-results}

The three cases with different priors are investigated with simulated evidence corresponding to the ``1994 discharge'' scenario in \cref{tab_reactor_history}.
The burnup input values are the centers of the burnup intervals and cooling time inputs are calculated with a discharge time in the center of the given operating window.
These input values are the ``true'' parameters of the five operating cycles that are to be reconstructed.
The corresponding batches are labeled in reverse alphabetical order, starting with the batch from 2016--2018 (Batch A) and ending with the batch from the production campaign from 1986--1994 (Batch E).

\cref{fig:1994_scenario} shows the posterior distributions obtained with the \emph{realistic priors}.
Only the burnup posterior of batch E somewhat resembles a normal distribution centered around the true parameter value.
The burnup posterior of batch A shows a small peak around the true value, while the remaining batches resemble uniform distributions.
The cooling time posteriors of each batch appear approximately uniformly distributed.

In contrast, the posterior distributions obtained with the \emph{wide priors} (shown in \cref{fig:full_hist_5_batch_wide_priors}) have clear peaks in the burnup of all five batches.
Furthermore, these peaks are concentrated on the true parameter values, indicating that the method is, in principle, sensitive to burnup, and can discriminate the different values of each batch.
The posteriors for cooling time still look like uniform or linear distributions in the given prior interval.
This suggests that the selected isotope ratios are not sensitive enough for the cooling times considered here.

The sensitivity to burnup is also visible in the third case, that is, the \emph{wrong priors}.
The posteriors obtained with these priors are shown in \cref{fig:full_hist_5_batch_wrong_priors}.
The posterior burnup distributions are not uniform anymore.
Instead, they are clearly skewed towards the lower limit, indicating where the true value is.
This suggests that the reconstruction process is well-suited to reject incorrect priors (i.e., false declarations) and identify the direction of the true value.
The reconstruction of the cooling time is not affected by this and resembles the results in the \emph{realistic priors} case.

\subsection{Model comparison}

We investigate the potential to distinguish between the two discharge scenarios with two inference models that differ in their number of cycles ($N_{c}$ in \cref{eq:ratio-model}).
The \emph{1994 model} assumes five reactor cycles and the \emph{pre-1994 model} assumes six cycles.
Simulated evidence is generated for both the ``1994 discharge'' scenario and the ``pre-1994 discharge'' scenario.
Both inference models are applied to to each test scenarios and ranked according to their LOO-CV weights.

\begin{figure}
    \centering
    \includegraphics[width=\linewidth]{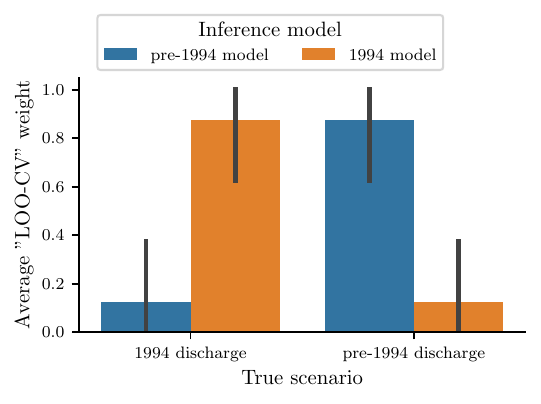}
    \caption{
        Scenario comparison with LOO-CV.
        The x-axis represents two scenarios with distinct sets of simulated evidence.
        Based on the LOO-CV, each model is assigned one weight per isotopic ratio.
        The y-axis shows the average weight assigned to each of the two inference models per scenario.
    }
    \label{fig:loo-results}
\end{figure}

\cref{fig:loo-results} shows the results of the two model comparison scenarios.
In the ``1994 discharge'' scenario, the \emph{1994 model} is ranked highest, while the \emph{pre-1994 model} is ranked highest in the ``pre-1994 discharge'' scenario.
Thus, in both scenarios, the model matching the number of cycles in the test scenario is preferred.

 \hypertarget{discussion}{\section{Discussion}\label[sec]{discussion}}

\begin{figure*}
    \centering
    \includegraphics[width=\textwidth]{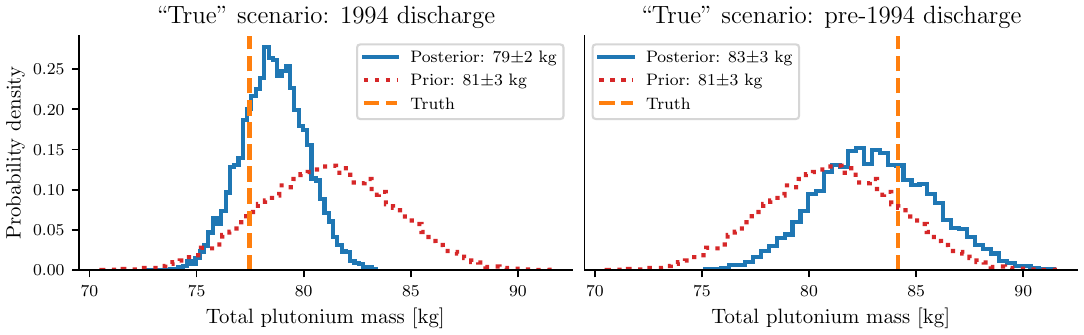}
    \caption{
        Plutonium production estimates derived from the prior and posterior distributions.
        The prior plutonium estimate is computed with equal weights for the priors of both the 1994 discharge and the pre-1994 discharge scenario.
        The posterior plutonium estimate is computed with the weights from the LOO-CV comparison of the two inference models.
        The left graph shows the posterior corresponding to a ``true'' 1994 discharge scenario and the graph on the right shows the posterior corresponding to a ``true'' pre-1994 discharge scenario.
        The x-axis show the total amount of plutonium produced over the complete reactor history, which are obtained by scaling the plutonium density in the simulated infinite lattice model with the total fuel volume of the reactor.
    }
    \label{fig:pu_estimate}
\end{figure*}

We have found that BRAM is in general suited to reconstruct the operational history for the 5MWe reactor at Yongbyon.
Using all of the available information as prior knowledge, the reconstruction does not directly yield more information, as the posterior distribution in \cref{fig:1994_scenario} shows uniform distributions for most variables.
However, the analysis with wider prior distribution (\cref{fig:full_hist_5_batch_wide_priors}) indicates that the method is indeed sensitive to the variables of interest.
Therefore, the results in \cref{fig:1994_scenario} are a powerful indication that the sampled evidence is consistent with the prior information.
It is suited to build confidence whether a declaration provided to a hypothetical inspecting entity is credible.

By modeling a case with incorrect prior limits, we showed that it is possible to detect incorrect assumptions or declarations.
Confronted with a potential scenario like this, where the distributions show the behavior displayed in \cref{fig:full_hist_5_batch_wrong_priors}, inspectors could demand explanations by the inspected state or request additional information to clarify the inconsistency.
Reanalyzing the case with updated, correct prior limits should lead to the posteriors such as shown in \cref{fig:full_hist_5_batch_wide_priors} and indicate the correct parameter values.

The capability to select between reactor histories with a different number of operating cycles is especially relevant for the present scenario, as there is some doubt in the open literature about whether any fuel was discharged from the reactor before 1994.
Although our example merely distinguishes between a full-core discharge and  no discharge before 1994, the model could be adapted to consider partial discharge scenarios as well.

\subsection{Implications for plutonium stockpile estimations}
\label{subsecplutonium}
The goal of reconstructing the operating history of a reactor is to verify the amount of plutonium that was produced in it.
A simple plutonium estimate can be obtained by inserting the posterior burnup samples into a model that calculates the plutonium mass density in the fuel.
To estimate the total amount of plutonium, the mass densities are scaled with the estimated fuel volume of the core and summing over the operating cycles.

\cref{fig:pu_estimate} illustrates such plutonium production estimates for the two model comparison scenarios in \cref{fig:loo-results}.
Since these scenarios consider two inference models, the plutonium estimates of each model are added together with weights according to the average weight calculated by LOO-CV.
The graphs also show the simulated ``true'' plutonium production for each scenario and a prior plutonium estimate derived from random samples drawn from the prior burnup distributions.
The true values lie well within the range covered by the posterior distributions, although the peaks are slightly shifted.
This example demonstrates that the LOO-CV-based model comparison approach can be useful to incorporate the possibility of alterative reactor histories into the plutonium estimates and potentially reduce associated uncertainties.

\subsection{The integrated approach to nuclear archaeology}\label{subsec:integrated-approach}

It is clear that the output of any single scientific or technical method will be accompanied by uncertainties.
To mitigate related doubts, it can be useful to employ an integrated approach to nuclear archaeology \cite{gottscheNuclearArchaeologyAssess2024}.
With this approach, the results of multiple methods and sources of evidence are compared and evaluated for consistency.
For example, the Graphite Isotope Ratio Method (GIRM) \cite{geshGraphiteIsotopeRatio2004} has also been proposed as a useful method to verify North Korea's past plutonium production \cite{Kütt}.
Both GRIM and BRAM could be used independently to estimate the fissile material stockpile.
If both results are consistent with the declaration, the declaration becomes more credible.
 \hypertarget{conclusion}{\section{Conclusion}\label[sec]{conclusion}}

In this simulated study, we demonstrate several promising aspects of the Bayesian Reprocessing Waste Analysis Method that underscore its potential to support verification efforts of future a North Korean fissile material declaration.
The primary utility lies in reconstructing the operating history of the \vmwe reactor in Yongbyon, North Korea, by inferring the burnup and cooling time of individual batches of spent fuel.
Using model comparison methods, the framework can distinguish different reactor histories and potentially provide new information on the early period, where there is a lack of open-source information.
The model weights derived with these methods can be used to calculate a weighted plutonium estimate that reflects the probability of two potential reactor histories.

Furthermore, the posterior distributions of the parameters of interest can be used to detect inconsistencies in the prior information.
If the application were designed to use a baseline declaration to define priors for the inference framework, such inconsistencies could indicate a false declaration.

Although these results are purely simulation based, they underscore the potential value of analyzing the reprocessing waste of the 5\,MWe reactor with a Bayesian inference framework to verify a future North Korean fissile material declaration.
Future research should focus on validating the methodology in experimental settings with known reactor histories to better understand how these results translate to real-world scenarios. 
\section*{Acknowledgments}
The authors gratefully acknowledge the funding provided by the VolkswagenStiftung with the Freigeist Fellowship (grant number 92 966-2) and the funding provided by the German Federal Ministry of Education and Research (BMBF) under grant number 01UG2210A (VeSPoTec).
The authors also gratefully acknowledge the computing time provided to them at the NHR Center NHR4CES at RWTH Aachen University (project number p0020230).
This is funded by the Federal Ministry of Education and Research, and the state governments participating on the basis of the resolutions of the GWK for national high performance computing at universities (www.nhr-verein.de/unsere-partner). \section*{CRediT author statement}
\textbf{Benjamin Jung:} Conceptualization, Methodology, Software, Data Curation, Investigation, Visualization, Writing - Original Draft
\textbf{Johannes Bosse:} Methodology, Software, Data Curation, Investigation, Writing - Original Draft 
\textbf{Malte G\"ottsche:}: Conceptualization, Writing - Review \& Editing, Supervision, Funding acquisition 
\bibliography{manuscript.bbl}
\end{document}